# UNIVERSALITY IN HADRON PRODUCTION IN ELECTRON-POSITRON, LEPTON-HADRON AND HADRON-HADRON REACTIONS

GEORGE D. LAFFERTY

*Department of Physics and Astronomy, The University of Manchester,*
*Manchester M13 9PL, UK*
*E-mail: George.Lafferty@man.ac.uk*

According to the concept of universality in hadron production, the basic mechanisms of hadron formation are the same in all high-energy $e^+e^-$, $l$h and hh reactions, with differences in the composition of final-state particle types being due only to differences in initial parton flavours and configurations. This concept is discussed in the light of recent data and phenomenology.

## 1 Introduction

Universality says that the mechanisms of hadron production are the same in all high-energy processes in which most of the final state hadrons arise from hard or semi-hard scattering of partons. Differences in the detailed composition of the final states are assumed to be due to differences in the flavours and configurations of the initial partons.

Perturbative quantum chromodynamics (pQCD) has been rigorously tested in many high-energy experiments, and the theory is now well established as the one describing the strong interactions of quarks and gluons. However, while many aspects of the experimental data, such as scaling violations, event shapes, jet rates and so on, are amenable to description in terms of pQCD, the detailed mechanisms of hadron formation are much less well-understood. These processes occur at the scale of the hadron masses and are essentially non-perturbative in nature. For studies of the production of hadrons in high-energy reactions we therefore have to rely on models to help us understand what is going on. A number of models exist, some to a greater or lesser extent built around features of QCD, while others are purely phenomenological.

## 2 Testing universality

To test the idea of universality in practice, one first has to understand the configuration and flavours of the scattering partons, in order to compare different types of processes. It helps also to have a theory or model that turns the partonic systems into hadrons. This is reasonably straightforward only in the case of $e^+e^-$ annihilation below the $W^+W^-$ threshold, shown schematically in Fig. 1a. Here, there are no hadrons in the initial state, and therefore no remnants of the initial state to worry about in the final state. The progenitor of the final-state consists, to first order, in a single quark-antiquark pair kicked out of the vacuum by a virtual photon or a Z-boson. As will be discussed below, there are several good models for the hadronization.

A schematic of the first-order (Born) term for lepton-hadron scattering is shown in Fig. 1b. Here things are manifestly more difficult, even for this lowest-order case, since one has to take into account the remnants the initial-state hadron. The difficulties are further compounded in the case of hadron-hadron scattering (Fig.1c) , where there are two hadrons in the initial state and so two sets of hadron remnants in the final state.

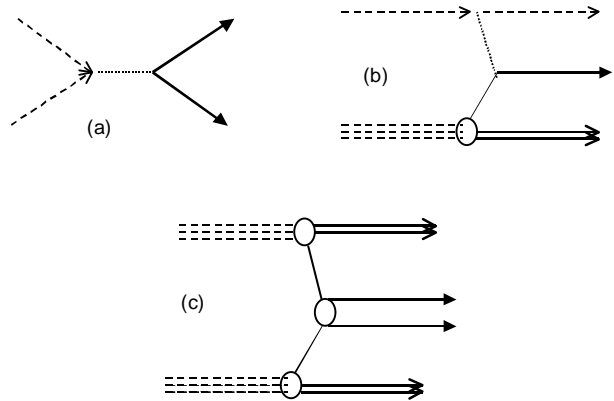

**Figure 1**. Simple schematic illustrations of three high-energy processes, in which outgoing partons are shown as solid arrows: (a) $e^+e^-$ annihilation — to first order the hadronic final state is initiated by a uniquely simple $q\bar{q}$ system; (b) the first-order (Born) term in deep-inelastic lepton-hadron scattering; (c) hadron-hadron scattering.

*2.1   $e^+e^-$ annihilation*

In the light of the above, it clearly makes sense initially to obtain as much information as possible about the process of hadron formation (hadronization) from the $e^+e^-$ annihilation data, and to develop and tune models to fit this simplest case. These models could then be applied, suitably amended to take account of different initial-state configurations, to other processes. One example of this practice is seen in studies of fully hadronic final states from $W^+W^-$ production at LEP 2. A detailed understanding of hadron production from single $q\bar{q}$ pairs, obtained with the LEP 1 data, helps in the analysis of the W$^+$W$^-$ data, where there are two q$\bar{\text{q}}$ pairs, one from each W decay. An important question in this particular case is whether the two q$\bar{\text{q}}$ systems hadronize independently or whether there is "colour-reconnection" between them. This question has a bearing on measurements of the W mass in these types of events, since the hadronization uncertainties are a major source of systematic error.

*2.2  LEP 1 data samples*

For studies of hadron formation in the simplest case, the best data we have are the Z decay data [1] from LEP 1. These data were taken between the years 1989 and 1995 over a range of center-of-mass energies from 88 to 94 GeV. Each of the four experiments, ALEPH, DELPHI, OPAL and L3 accumulated an integrated luminosity of about 170 pb$^{-1}$ (which seemed a lot at the time for an electron-positron collider; the machines feeding BABAR and BELLE can now deliver this amount in a single day, but one should not of course take such a naïve comparison too far.) These data provided good (but, of course, as in the nature of physics, never quite good enough) event statistics — approximately 4.3 million hadronic Z decays in each of the four detectors, which have been used to make a large number of measurements of the features of inclusive identified hadron production.[2]

*2.3  The "standard model" for* e$^+$e$^- \to$ Z$^0 \to$ q$\bar{\text{q}} \to$ hadrons

The process e$^+$e$^- \to$ Z$^0 \to$ q$\bar{\text{q}} \to$ hadrons is illustrated in Fig. 2, where five distinct phases (though not in the thermodynamic sense) are indicated. The electroweak standard model describes the initial production of the perturbative q$\bar{\text{q}}$ pair. There follows the fragmentation phase, in which perturbative QCD can be used; it is in this part of the process that the global event properties, such as numbers of jets and event shapes are determined. Exact calculations to some order using QCD matrix elements, or Monte Carlo simulations using parton shower algorithms, successfully reproduce the important event characteristics determined in this phase. The parton shower may evolve until the energy scale gets down to ~O(1 GeV), where perturbation theory is no longer applicable, and hadrons begin to form. Next there is the hadronization phase, which is the least well-understood part of the entire process, and the one that is most relevant to testing the concept of universality in hadron production. Monte Carlo models such as JETSET (now known more generally as PYTHIA), based on the Lund QCD string model, and HERWIG, which uses the QCD cluster model, in general do rather well here. Finally, the decays of particles and resonances and the detection of the final-state particles are well understood.

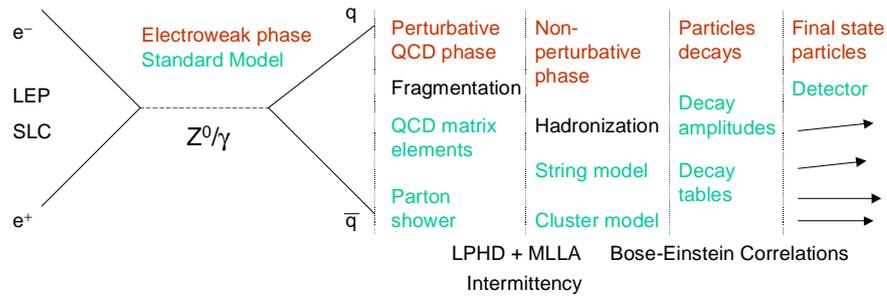

**Figure 2**. The "standard model" of hadron production in e⁺e⁻ annihilation at LEP 1 / SLC energies.

As Fig. 2 indicates, the separation of the entire process into separate phases is of course an approximation. Some phenomena that influence the nature of the final state are important over one or more phases; intermittency and Bose-Einstein correlations come into this category. Nevertheless, it is still useful to look at the process in this way, and to try to learn as much as possible about the hadronization phase. Indeed it is only by doing this that one can use Monte Carlo models at all.

There are some aspects of the initial $q\bar{q}$ state in Z decay that are important for the description of the final hadronic system:

- there are approximately 22% each of $d\bar{d}$, $s\bar{s}$ and $b\bar{b}$ pairs and 17% each of $u\bar{u}$ and $c\bar{c}$,
- the quarks are longitudinally polarized, with $P$(up-type) = –64% and $P$(down-type) = –94%,
- there is a forward-backward asymmetry of 10% for down-type and 7% for up-type quarks (polarized beams, as at the SLC, could produce much larger asymmetries).

*2.4 Local parton-hadron duality*

It is worth mentioning the modified leading log approximation (MLLA) and the concept of local parton-hadron duality (LPHD).[3] In MLLA the low-energy cut-off of the perturbative phase is pushed right down to the hadron mass scale. Then LPHD is basically the premise that the number and distribution of hadrons is the same as the number and distribution of partons at this cut-off scale. This approach seems to work well for quantities such as the center-of-mass energy dependence of the overall multiplicity and the momentum-distribution of the soft particles, but it says nothing about the detailed composition of particle types in the final state.

*2.5    The Breit frame in lepton-proton scattering*

Probably the next simplest configuration to study is provided by so-called Breit- frame analyses [4] in deep-inelastic electron-proton scattering. In this frame, shown schematically in Fig. 3, the quark struck by the space-like virtual photon recoils with equal and opposite momentum (for this reason, the Breit frame is occasionally called, drolly, the brick wall frame). In a quark-parton model picture, one then has two clearly separated event hemispheres, one corresponding to the current-region formed by the struck quark and the other to the target region containing the proton remnants. Jets of hadrons in the current region, arising primarily from hadronization of the struck quark, are comparable to single jets in two-jet events from $e^+e^-$ annihilation. Of course the situation is not quite so simple since there must be colour connections between the struck quark and the proton remnants, as there must also be between the q and $\bar{q}$ in $e^+e^-$ annihilation.

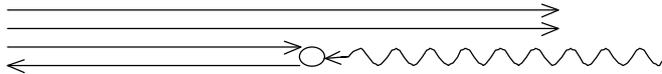

**Figure 3**. Schematic illustration of the Breit frame in deep-inelastic scattering.

*2.6    Models for the LEP 1 data*

A large number of measurements have been published from the LEP 1 data, including many measurements of inclusive rates and fragmentation functions (differential cross sections with respect to scaled momentum or energy) for single identified particles. The results can be compared to two classes of models: those, such as PYTHIA/JETSET [5] and HERWIG [6] which are complete Monte Carlo simulations that attempt to describe all features of the data; and a set of models which aim only to reproduce the measured inclusive rates.

2.6.1    Full Monte Carlo models

The Lund string model in JETSET has been reasonably successfully tuned to reproduce a large number of features of the data [1]. The program does have a large number of variable parameters, so to some extent it is not surprising that it can be made to fit the data. However, most of the parameters are physically motivated and the optimum set of values does provide useful information about the physics of hadron formation. In any case, Monte Carlo models are no good if they cannot reproduce the experimental data, because they are needed for studies of efficiencies, backgrounds, systematic errors etc. in many analyses.

The QCD cluster model of HERWIG is somewhat less successful than the Lund string model in fitting the Z decay data, particularly in the baryon sector. The cluster model has much less flexibility (i.e. fewer parameters to tune) than the string model, but this is not the only reason that it does less well. It is a fundamental premise of the model that the QCD clusters undergo isotropic decay into hadrons. Studies of the angular distributions of baryons from baryon-antibaryon pair systems [7], in the pair rest frame (specifically p$\bar{\text{p}}$ and $\Lambda\bar{\Lambda}$ pairs), show clear alignment along the jet axis, in excellent agreement with string model predictions. It is difficult to see how the cluster model could be modified to agree with these observations, without introducing an ad hoc fix that would simply make it more like a string model.

### 2.6.2 Models for inclusive rates of identified particles

There are some "magic formulae" in the literature [8], which have been developed to describe inclusive rates in the LEP 1 data. These formulae seem to work, but are generally phenomenological, without a great deal of physical underpinning.

In the string-based model approach of Yi-Jin Pei [9], inclusive identified hadron rates are defined solely in terms of three quantities: the hadron spin, $J$; a strangeness suppression factor dependent on the number of strange valence quarks, $N_S$, in the hadron; and the binding energy, $E_{\text{bind}} = M_{\text{h}} - \sum m_{\text{q}}$, of the hadron of mass $M_{\text{h}}$ composed of valence quarks of mass $m_{\text{q}}$. The total rate for a hadron of type h is then given by

$$<N_{\text{h}}> = C \frac{2J+1}{C_{\text{B}}} \left( \exp \frac{-\pi \left( m_{\text{s}}^2 - m_{\text{u}}^2 \right)}{\kappa} \right)^{N_{\text{s}}} \exp\left( \frac{-E_{\text{bind}}}{T} \right) \quad .$$

In this formula, $C$ is an overall normalization factor that depends on center-of-mass energy, $C_{\text{B}}$ is a relative normalization for baryons (equal to one for mesons), and $T$ is an effective hadronization temperature. The first exponential factor gives the strangeness suppression that arises in string fragmentation models due to the need for the produced quarks to tunnel out to the physical region ($\kappa$ is the string tension, typically 1 GeV fm$^{-1}$). The second exponential term is clearly a Boltzmann factor, although it isn't clear why, and if, thermal equilibrium is relevant to hadronization in such high-energy processes. This Pei model was first applied, rather successfully, to fit e$^+$e$^-$ annihilation data simultaneously at 10 GeV, 29-35 GeV and 91 GeV [9].

In another approach to the description of the $e^+e^-$ data, a thermodynamic model was developed by Becattini [10]. In this model, each jet is identified with a hadron gas phase (a fireball) in thermal and partial chemical equilibrium. The same comment as above applies here: the nature and timescale of the hadronization phase do not meet the usual criteria required for thermal equilibrium; nevertheless the model does reproduce the data remarkably well. There are three parameters: the temperature of the hadron gas; the volume of the gas; and a parameter for incomplete strange chemical equilibrium (equivalent to a strangeness suppression factor).

*2.7 Tests of universality*

After having been found to describe inclusive particle rates in $e^+e^-$ collision data over a range of center-of-mass energies, the Pei model was also used to describe data from inelastic pp and $p\bar{p}$ collisions [11]. Simply with the introduction of a plausible description of additional sea quark contributions in the hadron-hadron collisions, the model gives an excellent, simultaneous description of rates of identified light-flavoured hadron production in the $e^+e^-$, pp and $p\bar{p}$ data. In these fits, the same, universal value of the effective temperature, T = 270 MeV, applies for all of the data. The same set of parameters also describes heavy flavour hadron production in the $e^+e^-$ data.

The thermodynamical model of Becattini was also extended for hadron collision data, by Becattini and Heinz [12]. In simultaneous fits to $e^+e^-$, pp and $p\bar{p}$ data over a wide range of energies (from $\sqrt{s} = 19.4$ GeV up to $\sqrt{s} = 900$ GeV for the hadron collision data), the same temperature of around 165 MeV was found to fit for all of the data, indicating a universal freeze-out temperature in the hadronization. The hadron gas volume is approximately proportional to the total final-state multiplicity, rising from 10 fm$^3$ for $e^+e^-$ data at 29 – 35 GeV up to 20 fm$^3$ at LEP 1 energies.

**3 Conclusions**

It is impressive that both the Pei string-based model and the Becattini/Heinz thermodynamic model can describe $e^+e^-$, pp and $p\bar{p}$ data over a wide range of energies, with a universal set of parameters. Even though the models take quite different approaches to the physics, this must be taken as good evidence for universality in high-energy processes. However, it would be good to see these models tested also against ep data. There are relatively few measurements so far of identified particle production at Hera, but it is to be hoped that the forthcoming high-luminosity running should change that.

Total inclusive rates of identified particles are only one feature of the complex multiparticle final states. It is important to continue to develop full Monte Carlo models with a view to applying the same models to different physics processes in order to make more rigorous tests of universality. One vital step along the way is to obtain optimal tunings of the JETSET/PYTHIA model using the LEP 1 data (the simplest case). This has not yet been done. Maybe it is not possible, either because the parameter space is too large, or the model isn't good enough. But we need to continue working on it, and to apply the results to other high-energy reactions.

## 4  Acknowledgements

I would like to thank the organizers of the XXX International Symposium on Multiparticle Dynamics for a most informative, stimulating and enjoyable conference.

**References**

1. Knowles I.G. and Lafferty G.D., *J.Phys.G: Nucl.Part.Phys.* **23,** 731 (1997) .
2. Bohrer A., LEP-1, *Phys.Rep*. **291,** 107 (1997) .
3. See, for example, Dokshitzer Yu. L., Khoze V. A. and Troian S.I., *J.Phys.G*. **17,** 1585 (1991); Ochs W., Soft limits of multiparticle observables and parton hadron duality, e-Print Archive: hep-ph/9910319.
4. ZEUS Collaboration, Breitweg J. *et al*, *Eur.Phys.J.* C **11**, 251 (1999); H1 Collaboration, Fragmentation measurements in the Breit frame at Hera, Contributed to 29th International Conference on High-Energy Physics (ICHEP 98), Vancouver, Canada, 23-29 Jul 1998.
5. Sjöstrand T., PYTHIA 5.7 and JETSET 7.4: Physics and manual, LU-TP-95-20, CERN-TH-7112-93-REV, Aug 1995, e-Print Archive: hep-ph/9508391.
6. Marchesini G. *et al*, *Comput.Phys.Commun*. **67**, 451 (1992) .
7. ALEPH Collaboration, Barate R. *et a,*, *Phys.Rep*. **294** , 1 (1998).
8. Chliapnikov P.V. and Uvarov V.A., *Phys.Lett.* B **345**, 314 (1995) ; Chliapnikov P.V., *Phys.Lett.* B **462**, 341 (1999); Uvarov V., *Phys.Lett.* B **482**, 10 (2000) .
9. Yi-Jin Pei, *Z.Phys.* C **72**, 39 (1996) .
10. Becattini F., *Z.Phys.* C **69**, 485 (1996).
11. Yi-Jin Pei, e-Print Archive: hep-ph/9703243.
12. Becattini F. and Heinz U., *Z. Phys.* C **76**, 269 (1997).